\documentclass[prl, floatfix, superscriptaddress, longbibliography, reprint]{revtex4-2}
\usepackage[pdftex,plainpages=false,colorlinks=true,linkcolor=blue, citecolor=blue, urlcolor=blue]{hyperref}
\usepackage{amsfonts}
\usepackage{amsmath}
\usepackage{amssymb}
\usepackage{graphicx}
\usepackage{natbib}
\usepackage{color}
\usepackage{placeins}
\usepackage{physics}
\usepackage{orcidlink}

\begin{document}

\title{Correlation of maximum superconducting critical temperature with copper-oxygen energy distance and oxygen hole content in the Emery model}
\author{Eleanor M. O'Callaghan\, \orcidlink{0009-0006-0014-4153}}
\affiliation{Department of Physics, Royal Holloway, University of London, Egham, Surrey, UK, TW20 0EX}
\author{Nicolas Kowalski\, \orcidlink{0000-0002-7955-8394}}
\affiliation{D\'epartement de physique, Institut quantique \& RQMP, Universit\'e de Sherbrooke, Sherbrooke, Qu\'ebec, Canada J1K 2R1}
\author{A.-M. S. Tremblay\, \orcidlink{0000-0001-6932-8299}}
\affiliation{D\'epartement de physique, Institut quantique \& RQMP, Universit\'e de Sherbrooke, Sherbrooke, Qu\'ebec, Canada J1K 2R1}
\author{Giovanni Sordi\, \orcidlink{0000-0003-2481-7544}}
\email[corresponding author: ]{giovanni.sordi@rhul.ac.uk}
\affiliation{Department of Physics, Royal Holloway, University of London, Egham, Surrey, UK, TW20 0EX}
\date{\today}

\begin{abstract}
Identifying microscopic parameters that optimize the maximum superconducting critical temperature $T_c^{\rm max}$ in the canonical model of the copper-oxygen plane of cuprates, the Emery model, remains challenging. Using cellular dynamical mean-field theory at finite temperature, we find that for a fixed charge gap size in the parent charge-transfer insulating state, $T_c^{\rm max}$ unexpectedly increases with increasing the copper-oxygen energy distance, as this favors the transfer of electrons from oxygen to copper orbitals. We show that these findings emerge naturally in the Zaanen-Sawatzky-Allen scheme and capture observed trends in hole-doped cuprates. Overall, our study uncovers that $T_c^{\rm max}$ is optimized in the Emery model under three conditions: upon doping a charge-transfer insulator, close to the charge-transfer insulator to metal boundary, and deep into the charge-transfer regime. While the first two are well-established, uncovering the third is our core contribution. This finding indicates new paths for optimizing $T_c^{\rm max}$. 
\end{abstract}
 
\maketitle

{\it Introduction}-- 
Identifying the trends of the superconducting transition temperature $T_c$ in cuprates and connecting them to microscopic parameters of model Hamiltonians may help clarify the underlying mechanism of high-temperature superconductivity~\cite{Norman2011, keimerRev}. Known experimental trends of $T_c$ include (i) the dome-shape dependence of $T_c$ on doping~\cite{keimerRev}, (ii) the anticorrelation between the maximum superconducting critical temperature $T_c^{\rm max}$ and the size of the insulating charge gap from which superconductivity emerges on doping~\cite{Ruan:SciBull2016, Davis:PNAS2022, Wang:Science2023}, and (iii) the correlation between $T_c^{\rm max}$ and the transfer of electrons from oxygen to copper orbitals in the CuO$_2$ plane~\cite{Rybicki:NatComm2016, Jurkutat:PNAS2023}. 

Deciphering the microscopic mechanisms of these trends remains challenging~\cite{Norman2011, Kotliar:PhysC1988, Kotliar:PRB1988, Ohta:PRB1991, Feiner:PRB1992, Raimondi:PRB1996, Pavarini:PRL2001, Kent:PRB2008, ArrigoniCuO2, Weber2011, Lorenzo3band, Dash:PRB2019, Nicolas:PNAS2021, Mai:PRB2021, Mai:npj2021, Vucicevic:PRB2024, Cui:NatComm2025, BacqLabreuil:PRX2025, St-Cyr:2025, Vadnais:2026, Jacob:2026}. The third trend is the focus of the present work. The interplay between the second and the third trend is the focus of the companion article~\cite{Eleanor_long}. The first two experimental trends can be modeled at theory level by the single-band Hubbard model, but not the third. 

The single-band Hubbard model~\cite{Hubbard1963}, where electrons hop on a lattice and experience an onsite Coulomb repulsion, qualitatively accounts for the experimental variation of $T_c$ on doping and of $T_c^{\rm max}$ on the size of the Mott insulating gap~\cite{AMJulich, QinAnnuRev2022}. Decades of work show strong evidence that within the single-band Hubbard model $T_c$ is optimized under two conditions: {\it upon doping} the Mott insulating state that is realized {\it just above} the Mott insulator to metal boundary~\cite{Scalapino:1994, AMJulich, sshtSC, Gull:2013, LorenzoSC, QinAnnuRev2022}. This finding is confirmed by the more realistic three-band Emery model~\cite{Emery_1987, Varma_1987}, where electrons hop between two oxygen orbitals and one copper orbital and experience a local Coulomb repulsion on the copper orbital. Due to the Cu-O energy distance, the Emery model can be set in the charge-transfer regime, which is relevant for cuprates. 
Hence, the two conditions for optimizing $T_c$ in the single-band Hubbard model can be reformulated in the Emery model as follows: doping the charge-transfer insulating state and setting the system just above the charge-transfer insulator to metal boundary~\cite{Kotliar:IJMPB1991, Baumgartel:PRB1993, Scalapino:1994, Weber2011, Nicolas:PNAS2021}.

However, in the Emery model the same charge-transfer gap size, which quantifies how close the system is to the charge-transfer insulator to metal boundary, can be obtained for different values of the Cu-O energy distance. This raises the question of the dependence of $T_c$ on the Cu-O energy distance, {\it for a given value of the charge-transfer gap size in the undoped state}. This is the question we solve in this work within the Emery model. Note that this question cannot be addressed within the single-band Hubbard model, as the Cu-O energy distance is a parameter absent in that model. Unexpectedly, we found that $T_c^{\rm max}$  increases with increasing the Cu-O energy distance. This result opens up a new path for optimizing $T_c^{\rm max}$ and offers a prediction for recent proposals to realize the Emery model with ultracold atoms~\cite{Lange2026, McCabe2026}. 

The key to this unexpected finding is that the Cu-O energy distance controls the orbital occupancy and thus the distribution of the electrons in the CuO$_2$ plane. Therefore, addressing the dependence of $T_c$ on the Cu-O energy distance allows us to examine our results against the experimental correlation between $T_c^{\rm max}$ and the transfer of electrons from oxygen to copper orbitals in the CuO$_2$ plane~\cite{Rybicki:NatComm2016, Jurkutat:PNAS2023} (i.e. the third experimental trend mentioned above). We found that the Emery model, for a given value of the charge-transfer gap size in the undoped state, qualitatively captures the observed trend in hole doped cuprates, offering a theoretical framework for exploring and optimizing $T_c$.

{\it Model and Method}-- 
We study the Emery model~\cite{Emery_1987} 
\begin{align}
H & = \sum_{\mathbf{k} \sigma} C_{\mathbf{k} \sigma}^{\dagger}  \left[ \mathbf{h}_{0}(\mathbf{k}) -\mu \mathbf{I} \right]  C_{\mathbf{k} \sigma} 
+ U_d \sum_{\mathbf{R}_{i}} n_{d \mathbf{R}_{i} \uparrow} n_{d \mathbf{R}_{i} \downarrow} ,
\label{eq:EmeryModel}
\end{align}
where $C_{\mathbf{k} \sigma}^{\dagger}=( d_{\mathbf{k} \sigma}^{\dagger},  p_{x \mathbf{k}  \sigma}^{\dagger}, p_{y \mathbf{k} \sigma}^{\dagger})$, $C_{\mathbf{k} \sigma}=\left( d_{\mathbf{k} \sigma},  p_{x \mathbf{k} \sigma}, p_{y \mathbf{k} \sigma} \right)^{T}$, and $d_{\mathbf{k} \sigma}^{(\dagger)}$ and $p_{\alpha \mathbf{k}  \sigma}^{(\dagger)}$, with $\alpha=x,y$, destroy (create) an electron of wave vector $\mathbf{k}$ and spin $\sigma$ in the Cu $3d_{x^2-y^2}$ and O $2p_{\alpha}$ orbitals, respectively, $n_{d \mathbf{R}_{i} \sigma} = d_{ \mathbf{R}_{i} \sigma}^{\dagger} d_{ \mathbf{R}_{i} \sigma}$ is the number operator for the Cu orbital at site $\mathbf{R}_{i}$, $U_d$ is the onsite Coulomb repulsion between electrons on the Cu orbital, and $\mu$ is the chemical potential. Setting the Cu-Cu lattice distance to unity, the noninteracting part of the Hamiltonian is
\begin{align}
\mathbf{h}_0 (\mathbf{k}) & =
\left(
\begin{array}
[c]{ccc}
\epsilon_{d} & V_{dp_x} & V_{d p_y} \\
V^\dagger_{dp_x} & \tilde{\epsilon}_p + W_{p_x p_x} & W_{p_x p_y} \\
V^\dagger_{dp_y} & W^\dagger_{p_x p_y} & \tilde{\epsilon}_p + W_{p_y p_y} 
\end{array} \right)  , 
\label{eq:h0}
\end{align}
where $V_{dp_\alpha} = t_{pd}\left(  1-e^{-ik_{\alpha}}\right)$, $W_{p_\alpha p_\alpha} = 2t_{pp}^\prime \cos k_{\alpha}$, $W_{p_x p_y} = t_{pp}\left(  1-e^{ik_{x}}\right)  \left(  1-e^{-ik_{y}}\right)$. Here, $t_{pd}$ is the hopping amplitude between nearest neighbor Cu-O orbitals, $t_{pp}$ and $t_{pp}^\prime$ are the hopping amplitudes between nearest neighbor and next nearest neighbor O-O orbitals,  $\epsilon_{d}$ is the onsite energy of the Cu orbital, $\epsilon_{p} = \epsilon_{p_x} = \epsilon_{p_y}$ is the onsite energy of the O orbitals, which, following Ref.~\cite{AndersenLDA} we renormalize by $\tilde{\epsilon}_p = \epsilon_p -2t_{pp}$. Taking $t_{pp}=1$ as unit of energy and temperature (with $k_B=1$), we set $\epsilon_{d}=0$, $t_{pp}^\prime=1$, $t_{pd}=1.5$ (as in Refs.~\cite{Lorenzo3band, Nicolas:PNAS2021, GiovanniPRB2025, GemmaPRB2025}) and we vary  $\tilde{\epsilon}_p -\epsilon_d \in [2,10]$, $U_d \in [7,16]$, $T \in [1/50,1/40]$ and $\mu$ to study hole doping $\delta \in [0, 0.10]$. 
 
We solve this model at finite temperature both in the normal and in the $d$-wave superconducting state with the cellular extension~\cite{maier, kotliarRMP, tremblayR} of dynamical mean-field theory (CDMFT)~\cite{rmp}. Competition with other states is not considered. CDMFT maps the Emery model onto a cluster impurity model embedded in a self-consistent bath of noninteracting electrons. Here we use the minimal cluster that captures $d$-wave superconductivity, i.e. a plaquette formed by $N_d=4$ Cu sites and $N_p=8$ O sites. We solve the cluster quantum impurity problem with the hybridization expansion continuous-time quantum Monte Carlo method (CT-HYB)~\cite{millisRMP, Werner:2006, hauleCTQMC, patrickSkipList} in the segment representation. More details are described in Refs.~\cite{Lorenzo3band, Nicolas:PNAS2021, Nicolas:Master}.

\begin{figure}
\centering{
\includegraphics[width=1.0\linewidth]{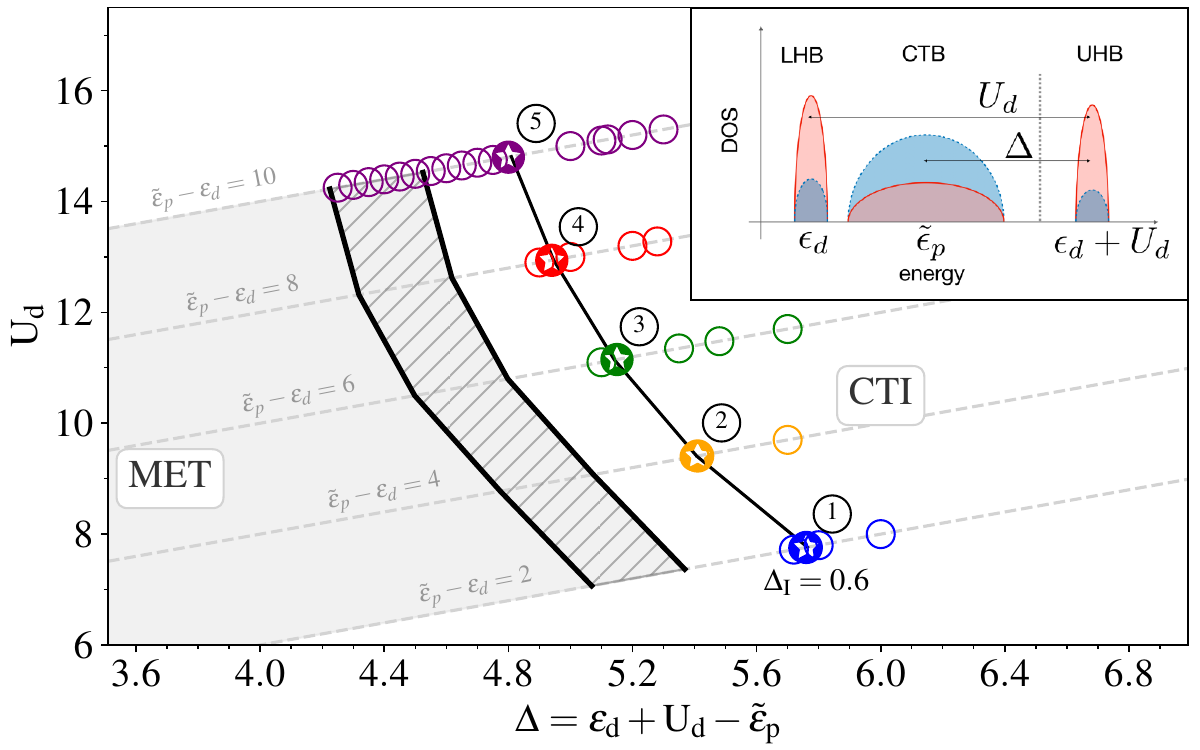}
}
\caption{Zaanen-Sawatzky-Allen diagram $U_d$ vs $\Delta = \epsilon_d +U_d - \tilde{\epsilon}_p$ for the Emery model. All data (circles) are at $n_{\rm tot}=5$, i.e. $\delta=0$, and $T=1/50$. The gray dashed lines indicate constant values of the bare Cu-O energy distance $\tilde{\epsilon}_p -\epsilon_d$. The shaded gray area denotes the metallic state (MET). The unshaded area denotes the charge-transfer insulating state (CTI). The metal to insulator boundary is first-order and accompanied by hysteresis (hatched region within thick black lines). The thin black line indicates the contour of constant charge gap size $\Delta_{\rm I} = 0.6$. Circles with star indicate the closest data points to that contour and are labeled by circled numbers. The superconducting state emerging from these charge-transfer insulators is analyzed in Figs.~\ref{fig:Tc} and \ref{fig:Tcstudy}. The inset shows a sketch of the Cu (red) and O (blue) partial density of states for a charge-transfer insulator. The charge-transfer band (CTB) is located between the lower and upper Hubbard band (LHB and UHB). The Fermi level (dotted line) lies between the charge-transfer band and the upper Hubbard band.
}
\label{fig:zsa}
\end{figure}

{\it Setting the model in the charge-transfer regime}-- 
Undoped cuprates are charge-transfer insulators~\cite{Emery_1987, ift, keimerRev}, where the insulating gap occurs between the charge-transfer band of mainly O character and the upper Hubbard band of mainly Cu character (see inset of Fig.~\ref{fig:zsa}). To set the Emery model in the charge-transfer regime, it is useful to build the Zaanen-Sawatzky-Allen diagram~\cite{zsa}, where the bare charge-transfer energy $\Delta= \epsilon_d +U_d - \tilde{\epsilon}_p$ (i.e. the energy distance between the upper Hubbard band and the onsite energy of the O orbitals, in the $t_{pd}\rightarrow 0$ limit) is on the $x$ axis and the interaction strength $U_d$ is on the $y$ axis. It is shown in Fig.~\ref{fig:zsa} for the low temperature $T=1/50$. Ref.~\cite{GemmaPRB2025} constructed the same diagram for $T=1/20$. To map out the $U_d -\Delta$ space, we conduct scans along lines of constant bare Cu-O energy distance $\tilde{\epsilon}_p -\epsilon_d$ (gray diagonal dashed lines). We choose $\tilde{\epsilon}_p -\epsilon_d \in [2,10]$ to ensure we span the charge-transfer regime defined by $U_d > \Delta$. If the Zaanen-Sawatzky-Allen diagram were extended to include the range of $\tilde{\epsilon}_p -\epsilon_d$ where $U_d < \Delta$, the Emery model would enter the Mott-Hubbard regime (see e.g. Ref.~\cite{GemmaPRB2025}).

The Zaanen-Sawatzky-Allen diagram in Fig.~\ref{fig:zsa} describes the Emery model at $n_{\rm tot}=5$, i.e. at one hole per CuO$_2$ unit cell. Open circles denote the data points explored in this work. For each data point, we compute the charge gap size $\Delta_{\rm I}$ by the width of the plateau at $n_{\rm tot}(\mu)=5$, which signals the incompressible correlated insulating phase. By evaluating $\Delta_{\rm I}$ as a function of $U_d$ at constant $\tilde{\epsilon}_p -\epsilon_d$, we can first determine the boundary between the charge-transfer insulating phase (where $\Delta_{\rm I}$ is nonzero) and the metallic phase (where $\Delta_{\rm I}$ is zero). Note that this boundary is first-order and accompanied by hysteresis (hatched region bounded by thick black lines), as shown in the companion article~\cite{Eleanor_long} (see also Ref.~\cite{Lorenzo3band} for $\tilde{\epsilon}_p -\epsilon_d=7$). Furthermore, we can construct lines of constant charge gap size $\Delta_{\rm I}$ in the $U_d -\Delta$ space (see companion article~\cite{Eleanor_long}). Specifically, in Fig.~\ref{fig:zsa} we show the line at $\Delta_{\rm I}=0.6$ (thin black line). 

\begin{figure*}
\centering{
\includegraphics[width=1.0\linewidth]{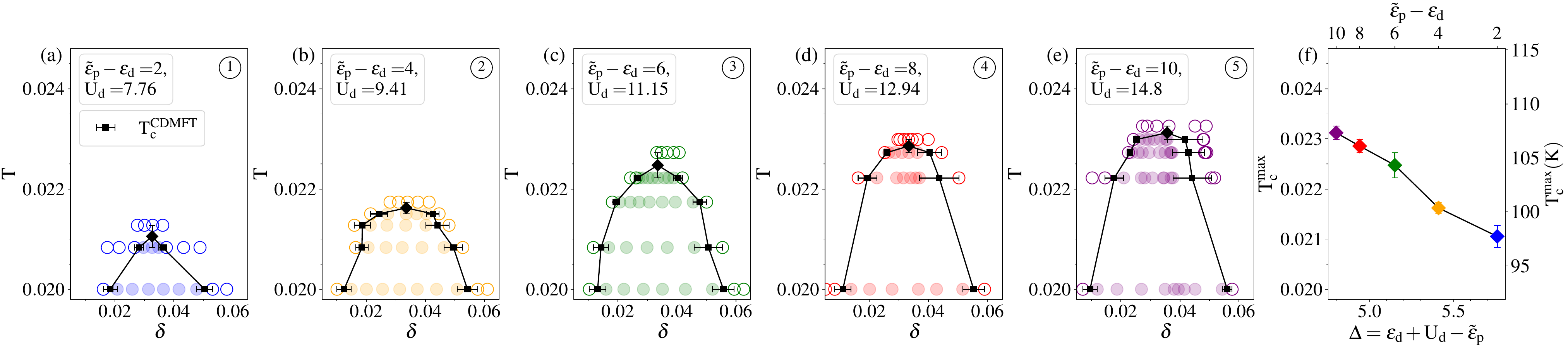}
}
\caption{(a)-(e) Superconducting temperature $T_c^{\rm CDMFT}$ (black line) versus hole doping $\delta=5-n_{\rm tot}$. $T_c^{\rm CDMFT}(\delta)$ is defined as the line where the superconducting order parameter changes from zero (open circles) to nonzero (shaded circles). The model parameters ($\tilde{\epsilon}_p -\epsilon_d$, $U_d$) are labeled and yield the same charge gap size $\Delta_{\rm I} =0.6$ in the parent ($\delta=0$) insulating state across all panels. These parameters are labeled by circled numbers that correspond to those in Fig.~\ref{fig:zsa}. (f) $T_c^{\rm max}$ versus bare charge-transfer energy $\Delta$, or $\tilde{\epsilon}_p -\epsilon_d$ (secondary $x$ axis), for the constant value  of charge gap size $\Delta_{\rm I} =0.6$ in the undoped state. Temperature is converted into Kelvin using $t_{pp}=0.4$~eV.
}
\label{fig:Tc}
\end{figure*}

{\it Superconducting temperature vs bare charge-transfer energy}-- 
Next, we study the superconducting state emerging from doping with holes the charge-transfer insulators found in the $U_d-\Delta$ space of Fig.~\ref{fig:zsa}. Hole doping is defined as $\delta=5-n_{\rm tot}$. 
We examine the superconducting state emerging from the charge-transfer insulators with the {\it same charge-transfer gap size} $\Delta_{\rm I}$ and different values of $\tilde{\epsilon}_p -\epsilon_d$ and $U_d$. This is a crucial point of our work. Specifically, we target the charge gap size $\Delta_{\rm I} = 0.6$, i.e. we study the superconducting state emerging from the parent charge-transfer insulating states labeled 1 to 5 in Fig.~\ref{fig:zsa}. We choose $\Delta_{\rm I} = 0.6$ because it is close to the metal-insulator boundary. Prior theory work~\cite{Nicolas:PNAS2021} has shown that the superconducting transition temperature decreases with increasing the distance from the metal-insulator boundary (see also companion paper~\cite{Eleanor_long}). This points to a link between superconductivity and superexchange mechanism. Ref.~\cite{Nicolas:PNAS2021} further supports this link by showing that superconducting pairing arises from low energy scales set by the superexchange $J$, and thus it is associated with short-range spin singlet correlations generated by $J$, similarly to the single-band Hubbard model~\cite{Kyung:2009, Eleanor_PRL2026}. 

We map out in Fig.~\ref{fig:Tc}(a)-(e) the superconducting temperature $T_c^{\rm CDMFT}$ vs doping $\delta$ emerging from the parent charge-transfer insulating states labeled 1 to 5 in Fig.~\ref{fig:zsa}.  We determine $T_c^{\rm CDMFT}$ as the temperature below which the superconducting order parameter is nonzero (see companion article~\cite{Eleanor_long}). Data points are marked with a shaded circle when the system is superconducting and an open circle when it is not. Despite in two dimensions thermal fluctuations preclude long-range order~\cite{MWtheorem}, $T_c^{\rm CDMFT}$ is a mean-field temperature neglecting Kosterlitz-Thouless physics and denotes when superconducting pairs develop within the cluster. 

In Fig.~\ref{fig:Tc}(a)-(e), $T_c^{\rm CDMFT}(\delta)$ increases from $\delta=0$, reaches a maximum $T_c^{\rm max}$ at finite doping and then decreases with further increasing $\delta$. As a result, the superconducting state shows a dome-like shape vs doping, similar to the experimental trends observed in cuprates~\cite{keimerRev}. 
With the same model and method used here, Refs.~\cite{Lorenzo3band, Nicolas:PNAS2021} showed a superconducting dome in the temperature-doping diagram for a given bare Cu-O energy distance in the range $\tilde{\epsilon}_p -\epsilon_d \in [6,7]$. Here, we map out the superconducting state vs doping for the wider range of  $\tilde{\epsilon}_p -\epsilon_d \in [2, 10]$. 

The first main result of this work is the dependence of $T_c^{\rm max}$ on the bare charge-transfer energy $\Delta$. Fig.~\ref{fig:Tc}(f) shows that for a constant charge-gap size $\Delta_{\rm I}$ in the parent state, $T_c^{\rm max}$ decreases with increasing $\Delta$. Equivalently, $T_c^{\rm max}$  decreases with decreasing the bare Cu-O energy distance $\tilde{\epsilon}_p -\epsilon_d$ (secondary $x$ axis in Fig.~\ref{fig:Tc}(f)), confirming the conjecture of Ref.~\cite{GemmaPRB2025} based on the normal state. This finding is compatible with the data in Refs.~\cite{Weber2011, Nicolas:PNAS2021} and challenges the prior thinking that $T_c^{\rm max}$ anticorrelates with the bare Cu-O energy distance $\tilde{\epsilon}_p -\epsilon_d$~\cite{Nicolas:PNAS2021}. Note that both the charge gap size $\Delta_{\rm I}$ of the parent state and the hopping amplitudes are kept fixed in our results.
 
It is convenient to map this finding onto the Zaanen-Sawatzky-Allen diagram $U_d - \Delta$ of Fig.~\ref{fig:zsa}. As we move along the black line at $\Delta_{\rm I}=0.6$ from bottom right to top left in Fig.~\ref{fig:zsa}, the doped model shows a higher $T_c^{\rm max}$. In other words, hole doping a system at point labeled $5$ in Fig.~\ref{fig:zsa} results in a higher $T_c^{\rm max}$ than that obtained by hole doping a system at the point labeled $1$. Physically, this means that systems with higher superconducting transition temperature lie {\it deep} in the charge-transfer regime of the Zaanen-Sawatzky-Allen diagram (i.e. in regions where $U_d \gg \Delta$ in Fig.~\ref{fig:zsa}), or equivalently far away from the Mott-Hubbard regime where $U_d < \Delta$. For a fixed charge gap size $\Delta_{\rm I}$, lowering the bare charge-transfer energy $\Delta$ (or equivalently, increasing the bare Cu-O energy distance $\tilde{\epsilon}_p -\epsilon_d$) results in a higher $T_c^{\rm max}$ in the doped model. 

In the companion article~\cite{Eleanor_long} we show that these findings are not restricted to $\Delta_{\rm I} =0.6$. By examining the interplay between $\Delta_{\rm I}$ and $\Delta$ on $T_c^{\rm max}$, we also show that systems with higher $T_c^{\rm max}$ lie in the Zaanen-Sawatzky-Allen diagram both deep in the charge-transfer regime (i.e. $U_d \gg \Delta$) {\it and} close to the metal-insulator boundary.

\begin{figure*}
\centering{
\includegraphics[width=1.0\linewidth]{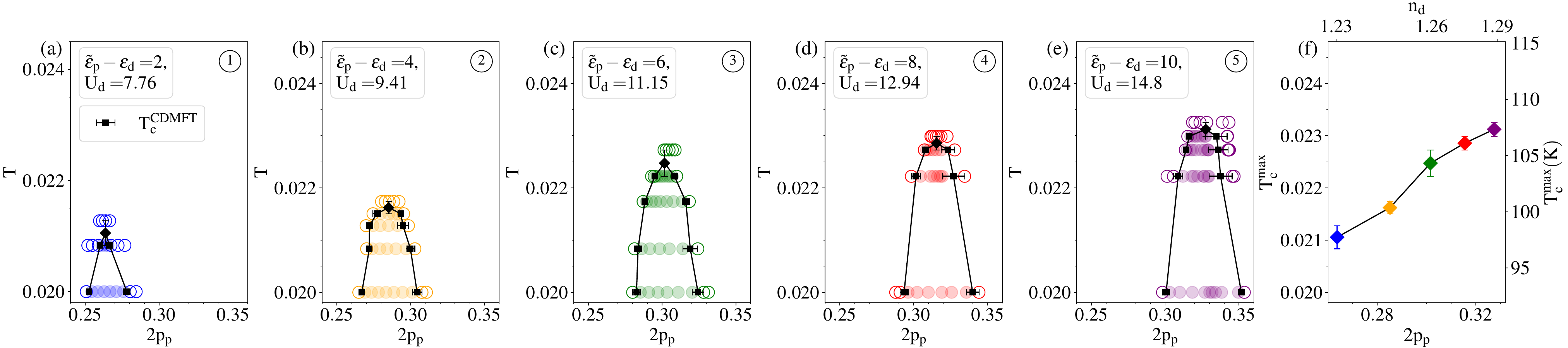}
}
\caption{(a)-(e) Same as Fig.~\ref{fig:Tc}, but using oxygen hole content $2p_p$ instead of hole doping $\delta$ on the $x$-axis. (f) $T_c^{\rm max}$ versus its corresponding oxygen hole content $2p_p$, or copper occupancy $n_d$ (secondary $x$ axis), for the constant charge gap size $\Delta_{\rm I} =0.6$ in the parent insulating state. Temperature is converted into Kelvin using $t_{pp}=0.4$~eV. 
}
\label{fig:Tcstudy}
\end{figure*}

{\it Superconducting temperature vs oxygen hole content}-- 
To understand the unexpected increase of $T_c^{\rm max}$ with decreasing bare charge-transfer energy $\Delta$ at a fixed charge gap size $\Delta_{\rm I}$ in Fig.~\ref{fig:Tc}(f), we follow the insight of the experimental Refs.~\cite{Rybicki:NatComm2016, Jurkutat:PNAS2023}. There, it is observed that the maximum superconducting temperature in cuprates is proportional to the oxygen hole content in the CuO$_2$ plane. The oxygen hole content is defined as $2p_p = 2(2-n_p)$, where $n_p$ is the O occupancy. Correspondingly, the copper hole content is defined as $p_d = 2-n_d$, where $n_d$ is the Cu occupancy.  Note that the partial occupancies $n_d$ and $n_p$ range from $0$ for an empty orbital to $2$ for a full orbital, so that the charge-transfer insulating phase occurs at $n_{\rm tot}=n_d+2n_p = 5$, or equivalently one hole per CuO$_2$ unit cell.  
Owing to the Cu-O hopping amplitude $t_{pd}$, the Cu occupancy $n_d$ is greater than $1$. Its deviation from unity is thus a measure of the mixed $d$-$p$ character of this hole. Physically, the hole that localizes in the charge-transfer insulator is shared between the Cu and O orbitals. 

To test with the Emery model the proportionality between $T_c^{\rm max}$ and $2p_p$ found in cuprates, first we follow the experimental Refs.~\cite{Rybicki:NatComm2016, Jurkutat:PNAS2023} and plot the superconducting critical temperature as a function of O hole content $2p_p$ rather than hole doping $\delta$. Fig.~\ref{fig:Tcstudy}(a)-(e) shows $T_c^{\rm CDMFT}$ vs $2p_p$ that emerges from the parent charge-transfer insulating states labeled 1 to 5 in Fig.~\ref{fig:zsa}, i.e. for fixed charge gap size $\Delta_{\rm I} = 0.6$ in the parent state and different values of the bare Cu-O energy distance $\tilde{\epsilon}_p -\epsilon_d$ and of interaction strength $U_d$. Note that Fig.~\ref{fig:Tcstudy}(a)-(e) and Fig.~\ref{fig:Tc}(a)-(e) contain the same data but with different $x$ axes. The superconducting state shows a dome-like shape as a function of O hole content, as observed in cuprates~\cite{Rybicki:NatComm2016, Jurkutat:PNAS2023}. The superconducting domes grow and move to higher O hole content upon increasing $\tilde{\epsilon}_p -\epsilon_d$ and $U_d$ (or equivalently, on decreasing $\Delta$), at a fixed charge gap size $\Delta_{\rm I}$. 

The second main finding of this work is the dependence of $T_c^{\rm max}$ on O hole content $2p_p$. Fig.~\ref{fig:Tcstudy}(f) collects the coordinates of $T_c^{\rm max}$ and  its corresponding O hole content for each panel. $T_c^{\rm max}$ increases with increasing $2p_p$. In terms of electrons rather than holes, this means that $T_c^{\rm max}$ increases with decreasing the O occupancy $n_p$ and correspondingly with increasing the Cu occupancy $n_d$ (secondary $x$ axis in Fig.~\ref{fig:Tcstudy}(f)). Physically, this indicates that increasing $T_c^{\rm max}$ correlates with the transfer of electrons from O to Cu orbitals. This finding captures the experimental trend on hole-doped cuprates revealed in Refs.~\cite{Rybicki:NatComm2016, Jurkutat:PNAS2023}. While in our work $2p_p$ varies within $[0.26,0.33]$, experimental values span a wider range, from nearly $0.25$ of La$_{2-x}$Sr$_x$CuO$_4$ (LSCO) to nearly $0.47$ of YBa$_2$Cu$_3$O$_{6+y}$ (YBCO), and to nearly $0.61$ of a Tl-based material.  Hence our model parameters quantitatively capture the values of LSCO only (see further discussion in the companion article~\cite{Eleanor_long}).

{\it Discussion}--
We can now understand the microscopic mechanism leading to the unexpected increase of $T_c^{\rm max}$ with decreasing bare charge-transfer energy $\Delta$ at a fixed charge gap size $\Delta_{\rm I}$ (Fig.~\ref{fig:Tc}(f)), and hence, why systems with higher superconducting transition temperature lie deep in the charge-transfer regime of the Zaanen-Sawatzky-Allen diagram (i.e. in regions of large $U_d$ and small $\Delta$ (large $\tilde{\epsilon}_p -\epsilon_d$) in Fig.~\ref{fig:zsa}). This is because these regions host a large O hole content. In terms of electrons, as the system goes deeper into the charge-transfer regime (while keeping fixed the charge gap size $\Delta_{\rm I}$), the mixed $d$-$p$ character of the doped holes is enhanced (see secondary $x$ axis in Fig.~\ref{fig:Tcstudy}(f)) due to the suppression of the electronic density on the O orbitals and its redistribution to the Cu orbitals. 

Therefore, this finding allows us to connect microscopic parameters of the Emery model ($U_d$, $\tilde{\epsilon}_p -\epsilon_d$) to measured trends in cuprates ($T_c^{\rm max}$ vs O hole content and charge-transfer gap size). Furthermore, our work shows that the Zaanen-Sawatzky-Allen diagram helps navigate the large parameter space of the Emery model: it allows us to intuitively understand how its parameters are related and to identify the increase of the O hole content as a microscopic mechanism controlling $T_c$, as found in experiments~\cite{Rybicki:NatComm2016, Jurkutat:PNAS2023}. From this framework, within the Emery model $T_c^{\rm max}$ is optimized under three conditions: upon doping a charge-transfer insulator, close to the charge-transfer insulator to metal boundary and deep into the charge-transfer regime (i.e. in the region of large $U_d$ and small $\Delta$, or equivalently large bare Cu-O energy distance), as this region displays large oxygen hole content. While the first two conditions were previously established~\cite{Kotliar:IJMPB1991, Baumgartel:PRB1993, Scalapino:1994, Nicolas:PNAS2021}, identifying the third condition is the key contribution of this  work. This finding opens up unexplored ways to optimize $T_c^{\rm max}$ and provides predictions for recent proposals to implement the Emery model with ultracold atoms in optical lattices~\cite{Lange2026, McCabe2026}.

\begin{acknowledgments}
We thank P. S\'emon for sharing his continuous-time quantum Monte Carlo code. We thank G. Reaney for discussions and work on a related project~\cite{GemmaPRB2025}. This work has been partially supported by the Canada First Research Excellence Fund. A.-M. S. T. benefits from \href{https://doi.org/10.69777/309032}{RQMP membership}. Simulations were performed on computers provided by the Canada Foundation for Innovation, Calcul Qu\'ebec, and Digital Research Alliance of Canada.
\end{acknowledgments}

\end{document}